\documentclass[twocolumn]{aastex701}
\usepackage{orcidlink}
\usepackage{amsmath}

\newcommand{\one}{\sc i\rm}

\shortauthors{Sextl \& Kudritzki}

\begin{document}

\title{The Hidden Story of Chemical Evolution in Local Star-Forming Nuclear Rings}

\correspondingauthor{Eva Sextl}
\email{sextl@usm.lmu.de}

\author{Eva Sextl \orcidlink{0009-0001-5618-4326}}
\affiliation{Universit\"ats-Sternwarte, Fakult\"at f\"ur Physik, Ludwig-Maximilians Universit\"at M\"unchen, Scheinerstr. 1, 81679 M\"unchen, Germany}
\email[]{sextl@usm.lmu.de}

\author{Rolf-Peter Kudritzki}
\affiliation{Institute for Astronomy, University of Hawaii, 2680 Woodlawn Drive, Honolulu, HI 96822, USA}
\affiliation{Universit\"ats-Sternwarte, Fakult\"at f\"ur Physik, Ludwig-Maximilians Universit\"at M\"unchen, Scheinerstr. 1, 81679 M\"unchen, Germany}
\email[]{kud@ifa.hawaii.edu}


\accepted{4th of December 2025}

\begin{abstract}
A VLT/MUSE population synthesis study of metallicities in the nuclear star-forming rings of four disk galaxies (NGC 613, NGC 1097, NGC 3351, NGC 7552) is presented. Disentangling the spectral contributions of young and old stellar populations, we find a large spread of ages and metallicities of the old stars in the nuclear rings. This indicates a persistent infall of metal-poor gas and ongoing episodic star formation over many gigayears. The young stars have metallicities a factor two to three higher than solar in all galaxies except NGC 3351, where the range is from half to twice solar. Previously reported detections of extremely metal poor regions at young stellar age on the rings of these four galaxies are a methodological artifact of the average over all stars, young and old. In addition, it is important to include contributions of very young stars ($<6$ Myr) in this environment. For each of the four galaxies, the extinction maps generated through our population synthesis analysis provide support for the infall scenario. They reveal dust lanes along the leading edges of the stellar bars, indicating the flow of interstellar material towards the circumnuclear zone. Prominent stellar clusters show little extinction, most likely because of the onset of stellar winds. Inside and on the nuclear rings, regions that are largely free of extinction are detected.
\end{abstract}



\section{Introduction}\label{sec:intro}
Nuclear star-forming rings and disks are prominent structures within the central kiloparsec of disk galaxies, where the gas density reaches levels sufficient to trigger intense localized star formation \citep{Knapen2005}. In many cases, they substantially contribute to the emission of the entire central galaxy region. Their formation is closely tied to the overall dynamical configuration of the host galaxy, especially the presence of large-scale stellar bars and other non-axisymmetric components that induce resonances (such as the Inner Lindblad Resonance) leading to gas accumulation in ring-like morphologies \citep{Athanassoula1992, Mazzuca2006,Verwilghen2024}. These dense gas reservoirs become sites of sustained starburst activity, and the ring often acts as a gas barrier that partially regulates the inward flow towards the nucleus. Over timescales of hundreds of millions of years, nuclear rings contribute to secular galactic evolution by building up stellar mass in the central regions and potentially influencing the growth of pseudo-bulges (inner disks) \citep{Kormendy2004,Sellwood2014}. \\
Despite their importance, the chemical evolution of such rings is still poorly understood. Gas-phase abundance measurements based on auroral lines are difficult in these regions because the relevant temperature-sensitive transitions are intrinsically faint and frequently overwhelmed by the bright stellar background \citep{Stasinska2005,Diaz2007,Bresolin2009}. The presence of nuclear activity further complicates the analysis by contaminating nebular emission lines with AGN-related radiation, which alters line ratios and can obscure the signatures of pure star formation \citep{Davies2014}. This effect is particularly relevant in the context of the well-established AGN-starburst connection, where inflowing gas concurrently fuels circumnuclear star formation and but also central supermassive black hole accretion, leading to intertwined episodes of starburst and AGN activity \citep{Clavijo2024}. \\
In the work presented here, we therefore do not focus on emission line analysis but instead on the young and old stellar components themselves. Unlike nebular line methods, which depend heavily on the detectability of temperature-sensitive transitions, stellar spectral features (Balmer lines, Calcium Triplet, iron lines, etc.) remain accessible even in regions with significant interstellar extinction. This allows for constraints on parameters such as ages, initial mass function (IMF) sampling, and metallicities through comparison with population synthesis models \citep{Leitherer1999, Bruzual2003}.  \\
In our population synthesis analysis of the integrated stellar spectra, we will employ the technique of full spectral fitting (FSF). FSF offers a newer and more powerful way to analyze nuclear SF rings by exploiting the entire stellar spectrum rather than isolated indices, allowing simultaneous constraints on stellar ages, metallicities, and kinematics \citep{Cid_Fernandes2005, Conroy2010}. By decomposing observed spectra into mixtures of simple stellar populations, FSF disentangles young starburst contributions from older bulge components and reveals star formation histories linked to bar-driven inflows. Despite the tremendous success of this technique, nuclear rings unfortunately show peculiar results. The flag-ship observational campaigns MUSE TIMER \citep{Gadotti2019} \& PHANGS-MUSE \citep{Emsellem2022} reported the presence of regions with apparently low metallicity despite exhibiting high H$\alpha$ luminosities indicating a very young population. The anomalous regions were persistent between different analysis methods and no adjustment of the fitting procedures was able to remove or reconcile them. In galaxies NGC 613, NGC 3351, NGC 1097, and NGC 7552, these signatures were especially pronounced \citep{Seidel2015, Bittner2020, Pessa2023, SilvaLima2025}, but other cases were also found \citep{Shimizu2019,Freitas20232, Robbins2025}. The sometimes extremely low ([Z]$<-0.5$) mean metallicities are truly peculiar given their circumnuclear location, where gas and subsequently stars are generally expected to be chemically enriched \citep{Perez2011, Cole2014}. \citet{Bittner2020} were cautious with physical interpretations and pointed out that their template set's lack of very young stellar populations (younger than $30$ Myr) might influence their results in these regions. \\
In this work, we show that for the complicated environment of central SF rings, mean light- or mass-weighted metallicity values obtained as averages over the total stellar population are not sufficient to characterize galactic evolution. We argue that we need another new approach besides 'light' and 'mass' weights to define metallicity. We introduce the 'physical' metallicity $Z_{\mathrm{phys}} = M_{metals}/M_{total}$ and show its usefulness in a 1-to-1 comparison with individual stellar probes in the disk of M83. We also disentangle the properties of young and old populations instead of discussing averages over all ages. With this new concept, we can tell another evolutionary story of some of the most prominent nuclear rings in the local universe. Finally, we emphasize the importance of an extensive age grid for spectral fitting templates in the complex environment of nuclear rings. Missing young stellar components in the fit introduces misleading results in nuclear ring studies.  

\begin{deluxetable}{lcccccccccc}
\label{galaxy_table}
\tabletypesize{\footnotesize}
\tablecaption{Properties of Sample Galaxies\label{tab:galaxies}}
\tablehead{
\colhead{Galaxy} &
\colhead{center $\alpha$} &
\colhead{center $\delta$} &
\colhead{stellar mass} &
\colhead{$i$ } &
\colhead{PA } &
\colhead{$D$} &
\colhead{spatial scale} &
\colhead{AGN ?} &
\colhead{Final S/N} \\
\nocolhead{} &
\colhead{(J2000)} &
\colhead{(J2000)} &
\colhead{$(\log M_{\odot}$)} &
\colhead{($^\circ$)} &
\colhead{($^\circ$)} &
\colhead{ (Mpc)} &
\colhead{(pc$/$arcsec)} &
\colhead{} &
\colhead{} &
\colhead{} }

\startdata
NGC 7552 & 349.044945 & -42.584962 & 10.52 & 14 & 54.9  & 17.2 &  83 & no & 200 \\
NGC 613  & 23.575714 & -29.418573 & 11.09 &  46 & 118 & 17.5 & 85 & yes & 130 \\
NGC 1097 & 41.578937 & -30.274717 & 11.24 & 46 &  130 & 14.5 & 70 & yes &  200  \\
NGC 3351 & 160.990618 & 11.703659 & 10.49  &  45  & 13  & 9.96 & 48 & no & 200  \\
\enddata
\tablecomments{Central right ascension and declination coordinates (centroid of the $3.6$ $\mu$m emission peak) as well as the stellar masses are taken from the Spitzer Survey of Stellar Structure in Galaxies (S4G; \citet{Sheth2010}). Position angles (PA) are measured east of north; $i$ is the inclination, $D$ are distances. The spatial scaling is calculated assuming the given distance. The primary literature references for the other quantities are \citet{Bittner2020} for NGC 7552, \citet{Sato2021} for NGC 613, \citet{Onishi2015} for NGC 1097, and \citet{Sun2024} for NGC 3351.}
\end{deluxetable}

\section{The galaxy sample}

Our sample consists of four nearby barred spiral galaxies that host some of the most prominent circumnuclear star-forming rings in the local universe: NGC 7552, NGC 613, NGC 1097, and NGC 3351. These systems are well-studied archetypes in which the interaction between bars and central star formation has been extensively documented. All four galaxies are included in the TIMER or PHANGS-MUSE survey and their ring like structures are kinematically well established. However, their subsequent analysis with respect to stellar metallicity, ages, and interstellar medium dust content presented significant challenges.\\
\textbf{NGC 3351}: This barred spiral galaxy hosts a well-defined circumnuclear star-forming ring at a radius of about $\sim300$ pc \citep{Swartz2006}. The ring is composed of regularly distributed HII-regions and massive young stellar clusters \citep{Colina1997,Bresolin2002}. The nucleus itself is dominated by an old stellar population, with no evidence of AGN activity \citep{Swartz2006,Pessa2023}. Gas inflow along the bar appears to efficiently feed the ring, while leaving the very center comparatively quiescent. The regularity and isolation of the nuclear ring make NGC 3351 a relatively 'clean' case for studies of ring star formation. \\
\textbf{NGC 1097}: \citet{Gadotti2019} noted that despite being the most massive galaxy in the present sample, it exhibits a prominent and extremely young starbursting ring of $\sim800$ pc radius. More than $300$ HII regions have been resolved in near-infrared imaging \citep{Prieto2005}. The extend of the ring is well-resolved by MUSE, allowing it to be identified with ease in this galaxy. Classified as LINER, NGC 1097 possesses a comparatively faint nucleus that we mask out. \\
\textbf{NGC 613}: This galaxy hosts a rather asymmetric nuclear ring with an $\sim400$ pc radius. \citet{Barroso2014} found an unusual large reservoir of molecular gas within $\sim100$pc. It harbors a radio jet, an outflow, and an AGN ionization cone in its nuclear region \citep{Silva2020}. The entire nuclear center shows complex gas kinematics influenced by AGN activity \citep{SilvaLima2025}. However, we do not detect broad-line region (BLR) features in the spectra. \\
\textbf{NGC 7552}: This ring is about $200$pc in radius. The central nucleus itself lacks starburst or Seyfert-like activity \citep{Forbes1994,Gadotti2019}. The ring is rich in molecular gas and hosts numerous massive young clusters, making it one of the most extreme star-forming nuclear rings in the nearby universe with a current star formation rate of $=10–15\, \mathrm{M}_{\odot} \,\mathrm{yr}^{-1}$, \citep{Pan2013}.

\begin{figure*} 
       \medskip
	\center \includegraphics[width=0.98\textwidth]{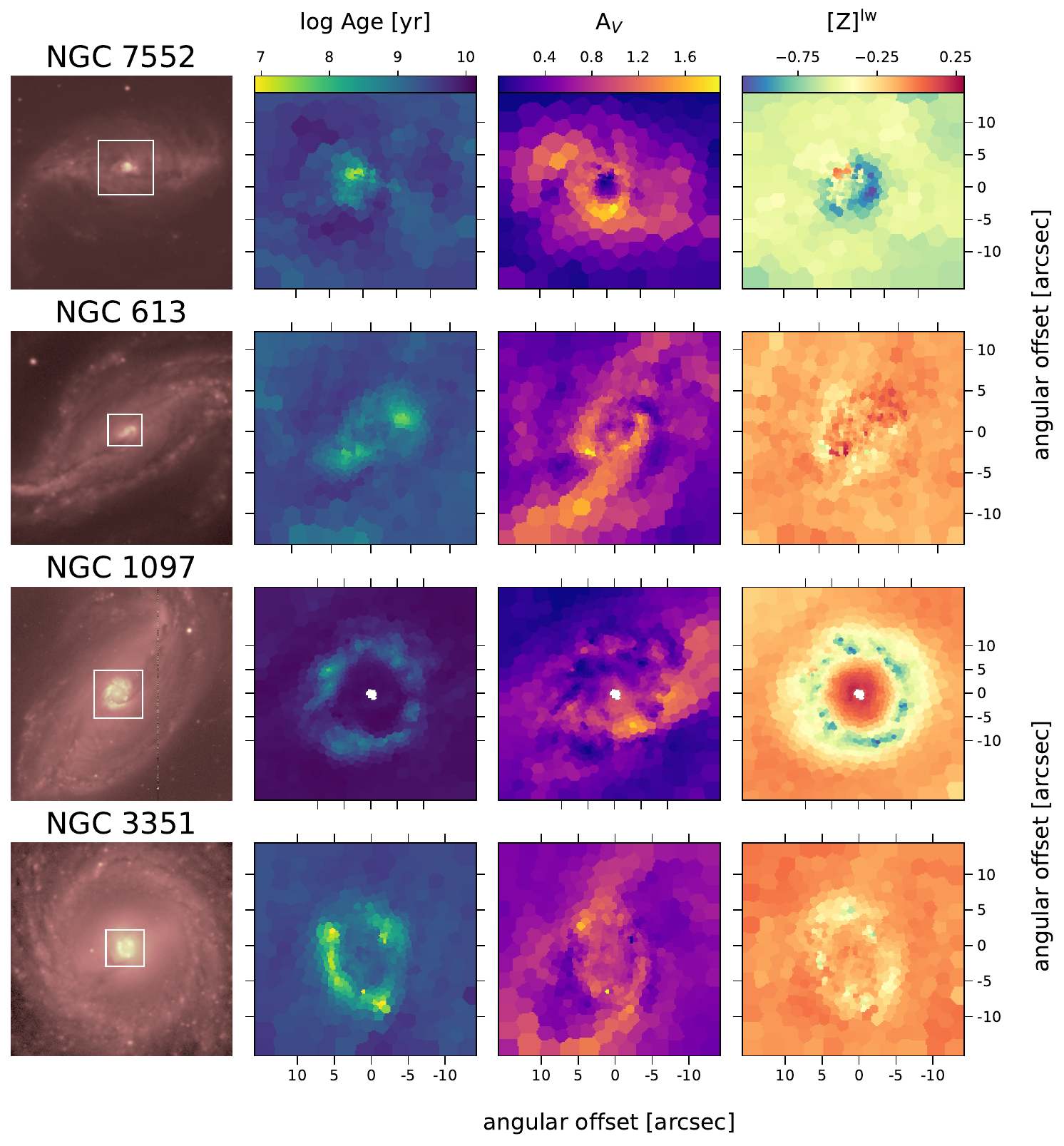}\medskip
	\caption{First column: B-Band images from the CTIO 1.5m telescope (NGC 7552), CTIO 0.9m telescope (NGC 613), du Pont 2.5m telescope (NGC 1097) and CTIO 1m telescope (NGC 3351). The MUSE FOV used for our FSF fit is marked in white. North is to the top, east is to the left. The following columns show the FOV with results from the FSF fit: mean light-weighted Age, visual extinction $A_V$ and light-weighted total metallicity [Z]$^{\mathrm{lw}}$. The color bar at the top holds for all the subplots in the column.}     \label{fig:Fig1_overview}
\end{figure*}

\begin{figure}
       \medskip
	\center \includegraphics[width=\linewidth]{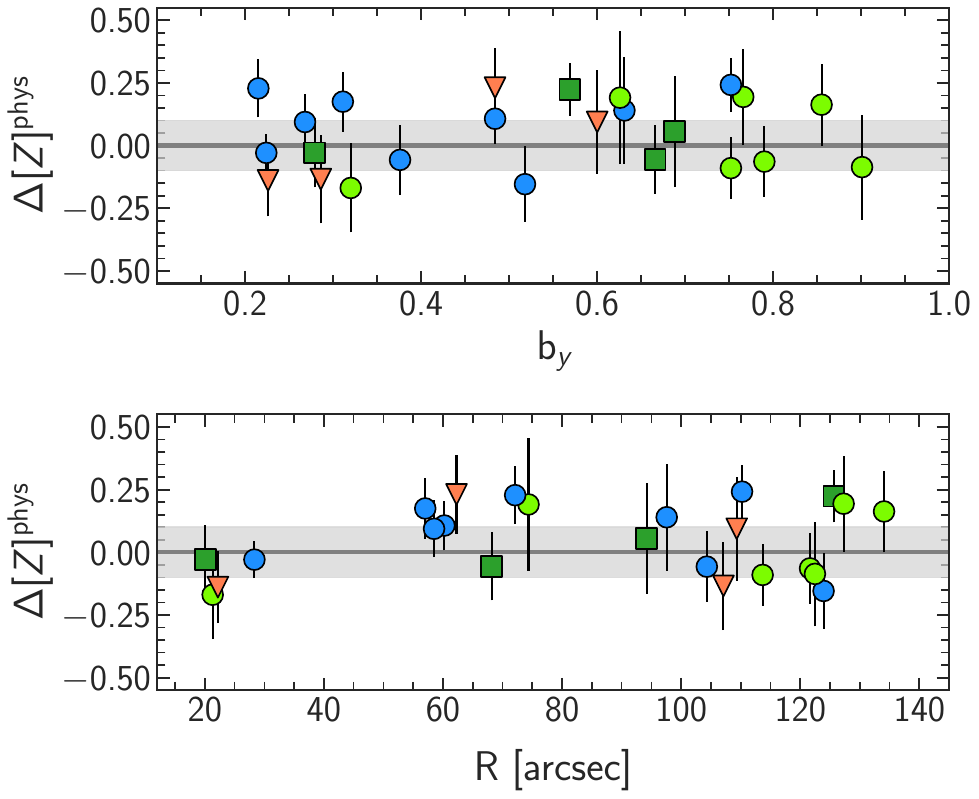}\medskip
	\caption{Metallicity difference $\Delta [Z]$ between individual stellar probes and population synthesis (see  Section \ref{M83_test}) as a function of the luminosity fraction b$_y$ of young stars (top) and the angular distance from the center (bottom). The different symbols represent differences with respect to BSG (blue circles), SSC (red triangles), YMC with optical analysis (dark green squares), YMC with UV analysis (light green circles). Errors result from the addition of stellar source and population synthesis errors in quadrature. The shaded gray strip indicates a difference $\leq 0.1$ dex and is added for orientation.}  \label{fig:M83_compare}
\end{figure}

\begin{figure}
	\center \includegraphics[width=\linewidth]{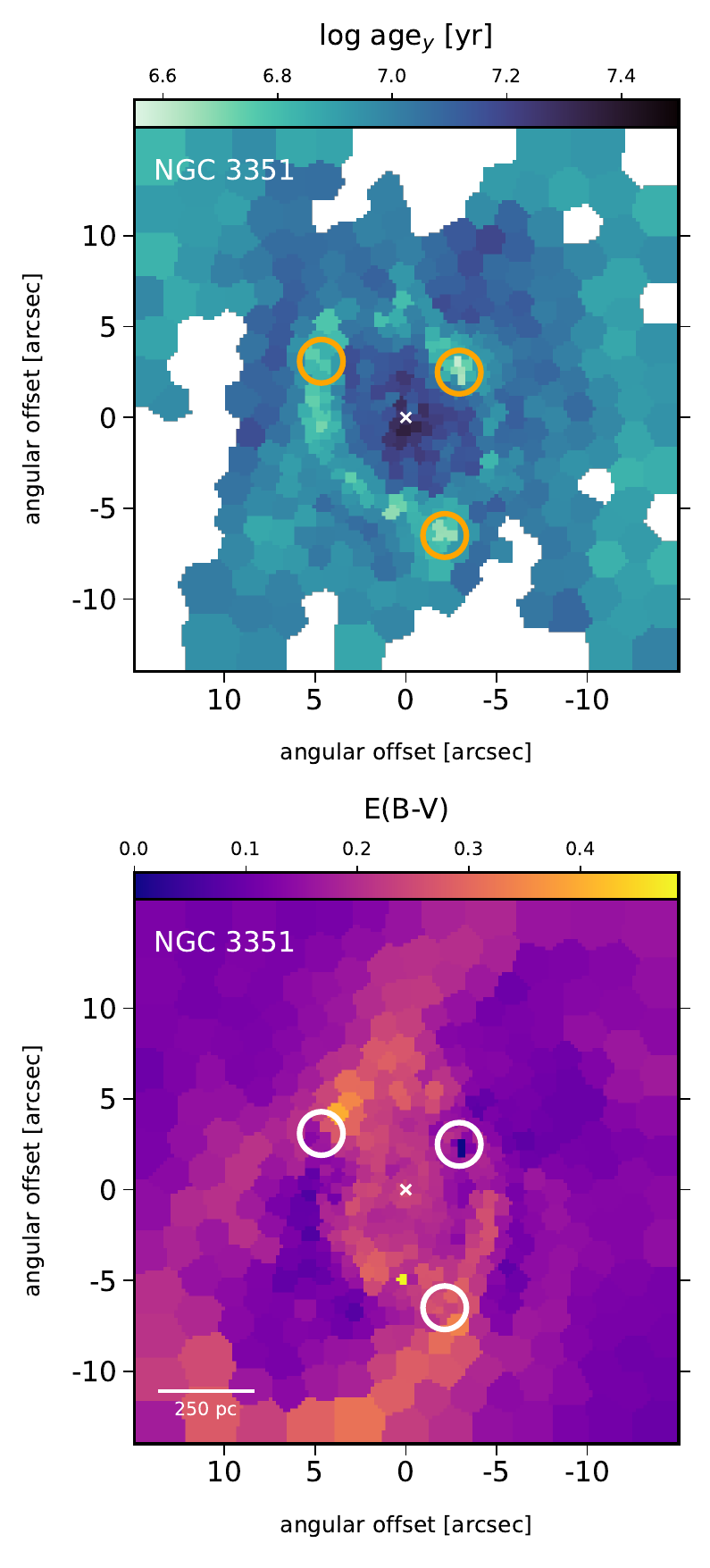}
	\caption{The central region of NGC 3351: (top) map of mean ages of the young stellar population; (bottom) interstellar reddening E(B-V). The circles indicate the locations of the three most prominent young stellar clusters highlighted in Fig.~24 of \citet{Emsellem2022} and discussed in the text.}  \label{fig:AgeY_map}
\end{figure}

\begin{figure}
       \medskip
    \center \includegraphics[width=0.99\linewidth]{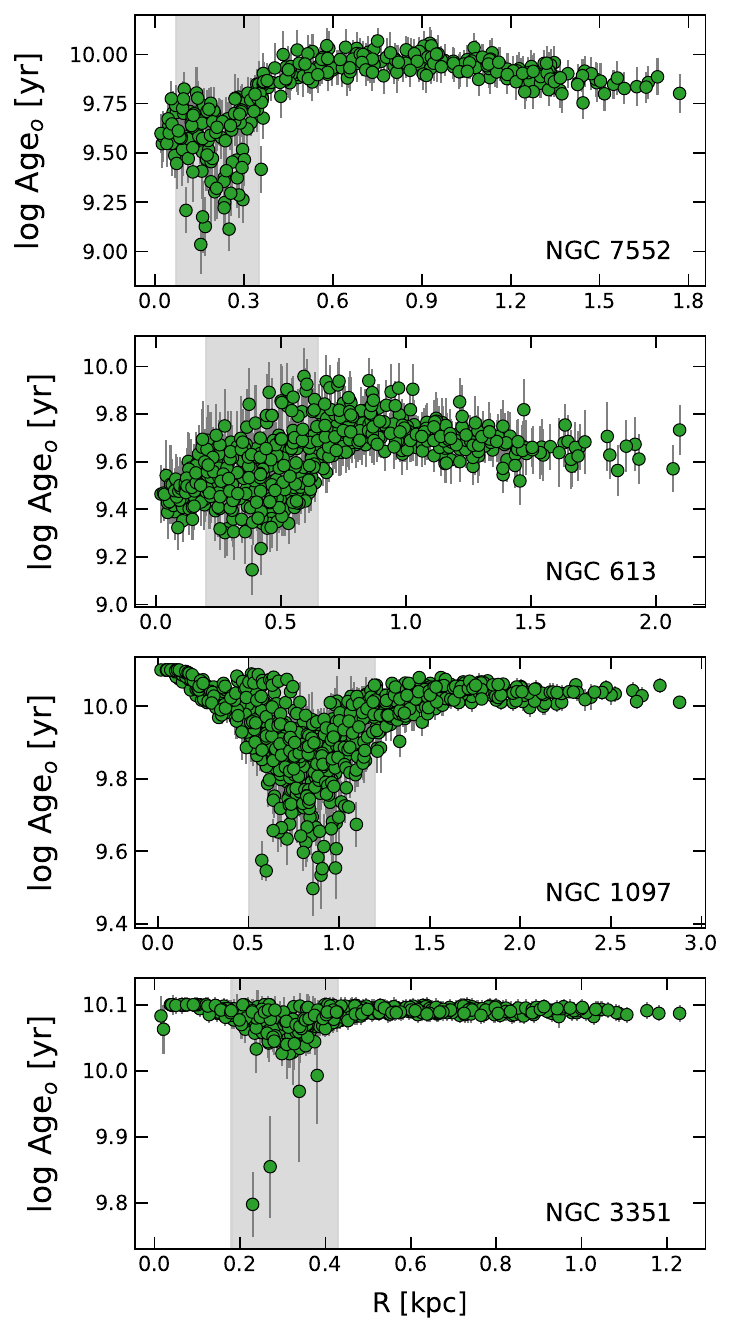}\medskip
	\caption{Mean ages of the old stellar population as a function of galactocentric distance. The nuclear star forming regions are indicated in gray. From top to bottom: NGC 7552, NGC 613, NGC 1097, NGC 3351.}  \label{fig:AgeO_grads}
\end{figure}

\section{Observations and data retrieval}

Our analysis relies primarily on observational data obtained with the Multi Unit Spectroscopic Explorer (MUSE, \citet{Bacon2010, Bacon2014}) on the Very Large Telescope (VLT) in Cerro Paranal, Chile. MUSE is an integral field spectrograph that provides spatially resolved spectra over a nominal wavelength range of 4800 to 9400 \AA\, in steps of 1.25 \AA~with a spectral resolution of $\sim$ 2.65 \AA~FWHM. The wide field covers a 60" by 60" field of view with 0.2" spatial sampling, enabling highly detailed mappings of galactic environments \citep{Weilbacher2020}.  \\
For all nuclear rings in our sample, we use the reduced IFU data cubes from the MUSE-DEEP program. At the VLT, observations are always organized into Observing Blocks (OBs), each representing a single pointing with a one-hour maximum exposure time due to operational constraints. Each OB produces a final data cube, referred to as an OB datacube, and these are released individually as part of the 'MUSE collection' in the Science Portal\footnote{\url{https://archive.eso.org/scienceportal/home}}. The MUSE-DEEP program combines multiple such OB datacubes for each target to create a single deep datacube with improved sensitivity. The data reduction in this case follows the standard MUSE pipeline optimized for deep field observations and includes bias subtraction, flat-fielding, wavelength and flux calibration, and sky subtraction. Figure~\ref{fig:Fig1_overview} shows B-band images for each galaxy in the sample. The subsequently used FOV is indicated in white. They are slightly smaller than the original cubes to better isolate the ring regions.  

\section{Data preparation \& Population synthesis technique}
As mentioned in the introduction, we make use of the widely adopted FSP technique with Simple Stellar Populations (SSPs) spectra in order to extract physical properties from the observed integrated light spectra. The method essentially models the observed spectrum as a linear superposition of SSP templates, each corresponding to a single-age, single-metallicity stellar population. The SSP sets are synthesized beforehand from theoretical stellar evolution isochrones and corresponding stellar spectra under an assumed initial mass function (IMF) (see Section \ref{subsec:SSPset}). \\
In mathematical turns, a model spectrum M$_{\lambda}$ is constructed from a collection of SSPs $f_{\lambda, i}$ with ages t$_{i}$ and metallicities [Z]$_i$ = log(Z$_{i}$/Z$_{\sun}$) as
\begin{equation}
  M_{\lambda} = D_{\lambda}(R_{V}, E(B-V)) \left[ \sum_{i=1}^{n_{SSP}} b_{i} f_{\lambda, i} (t_{i}, [Z]_{i})  \right].
\end{equation}  
The  fit coefficients b$_i$ are determined within the fitting procedure. We note that the SSP spectra (as well as the observed spectra) are normalized to unity in the range of 5500 to 5550 \AA, a wavelength regime without prominent emission or absorption features. Due to this normalization, the sum of all coefficients b$_i$ adds up to unity. The b$_i$ are luminosity weighted fit coefficients describing the contribution of the corresponding SSP to the integrated light at the wavelength of normalization (see \citealt{Sextl2023,Sextl2024,Sextl2025M83} for a detailed discussion). 

Interstellar dust along the line of sight is accounted for by the term D$_{\lambda}(E(B-V))$, whereby the colour excess E(B-V) is also fitted simultaneously. The overall shape of the attenuation curve is chosen in advance. We chose the prescription of \citet{Calzetti2000} with fixed $R_V=4.05$ due to the star-forming nature of nuclear rings. 

\subsection{Workflow}
In practical terms, the fitting process is performed with the pPXF algorithm \citep{Cappellari2004, Cappellari2023}. Inspired by the examples provided in pPXF’s GitHub repository\footnote{\url{https://github.com/micappe/ppxf_examples}. We also encourage the use the newer .dust-function for the extinction, not the now obsolete .reddening/ .gas-reddening keywords}, our approach follows several sequential steps which we have already successfully applied to TYPHOON IFU data in the past \citep{Sextl2025M83}.\\
When performing full-spectral fitting to reconstruct chemical evolution histories, it is a good practice not to work on the level of individual spectral pixels (spaxels), but to combine several to obtain a higher S/N spectrum. Voronoi binning is widely regarded as the standard approach for this purpose, and we applied it using the vorbin Python package \citep{Cappellari2003}. The ultimate S/N values at the wavelengths of normalization for each galaxy are listed in table~\ref{galaxy_table}. 

As a second step, we applied sigma-clipping to remove artifacts and emission lines in the Voronoi spectra, following Eq. (34) of \citet{Cappellari2023}. Data points that exceeded the relative error of 3$\sigma$ were excluded. In this process a multiplicative Legendre polynomial of degree $4$ was included to correct for low-order continuum flux mismatches. 

Third, the galaxy stellar kinematics were determined. Velocity and velocity dispersion were constrained in the pPXF main routine with an additive polynomial of degree 4. The high spatial and spectral resolution of MUSE would in principle also allow the extraction of higher-order moments (h$_3$ and h$_4$), but this lies beyond the scope of the present work. The implementation of a polynomial, which is added (not multiplied) to the spectrum, is described as best practice for kinematics in \citet{Cappellari2017}. It is especially necessary in (rare) cases when the available stellar templates differ from the true stellar population in the observed system. We also investigated the impact of the order $n$ of the polynomial and found it small (probably due to the generally superb S/N ratio).
 
The fitting then proceeded to the derivation of stellar population properties as a fourth step. During this stage, the kinematic parameters are fixed, and polynomial corrections are disabled to avoid degeneracies between continuum adjustments and population parameters. The parameters $b_i$ and $E(B-V)$ are determined simultaneously during this main fit. To obtain an estimate of the uncertainty in the fit, a wild bootstrapping procedure is applied $25$ times \citep{Davidson2008}. The residuals, which are the differences between the Voronoi spectrum and the best-fit model, are randomly multiplied by +1 or -1 to create a new set of perturbations. These perturbed residuals are added back to the original best-fit model to create many simulated spectra, which are then fitted again to see how the fit parameters $b_i$ and subsequently metallicities and ages may vary. This is a proper way to understand how uncertainties in the data (noise in the spectra) propagate into uncertainties in the discussed fitted parameters.

\subsection{Derivation of physical quantities}
Once the parameters $b_i$ are known, physically relevant quantities can be constructed, such as the mean light‑weighted age and the mean light‑weighted stellar metallicity $[Z]^{\mathrm{lw}}$ of the stellar population in the bin:
\begin{align}
 \log(t)^{\mathrm{lw}} &= \sum_{i} b_{i} \log(t_{i})/\sum_{i} b_{i}\\
 [Z]^{\mathrm{lw}} &=  \sum_{i} b_{i} \log(Z_{i}/Z_{\odot})/\sum_{i} b_{i}
\end{align}

These definitions follow directly from the spectral fit, but can be misleading, as young stellar populations can easily outshine older populations despite a much smaller total mass. The mass-weighted coefficients $\Tilde{b_{i}}$ can show a more nuanced picture and are obtained using the mass-to-light ratio $\gamma_{i}$ = M$_i$/L$_i(V)$ of each SSP isochrone as follows:
\begin{equation}
  \Tilde{b_{i}} = {\frac{b_{i}\gamma_{i}}{\sum_{i} b_{i}\gamma_{i}}}.
\end{equation}
The mass-weighted means for age and metallicity are then calculated in the same way as with equations (2) and (3) but using the coefficients $\Tilde{b_{i}}$ instead of $b_i$.

However, the definitions of light- or mass-weighted metallicities are misleading when compared with metallicities used in chemical evolution of galaxies or cosmological simulations. Here, metallicity $Z$ is the ratio of the mass of metals $M_{Z}$ confined in the stellar population divided by the total mass of stars $M$, $Z = M_{Z}/M$. Thus, following \citet{Sextl2025M83} we introduce a physical definition of metallicity: 
\begin{align} \label{physMetal}
     Z_{\mathrm{phys}} = {\frac{M_{Z}}{M}} &= \sum_{i} \Tilde{b_{i}}Z_{i} \\
     [Z]^{\mathrm{phys}}= \log(Z_{\mathrm{phys}}/Z_{\odot})
\end{align}
Note that this definition is \emph{different} from the 'mass-weight' normally found in the literature. We therefore call equation~\ref{physMetal} the 'physical metallicity' to distinguish the two quantities. Our choice is particularly relevant, as it allows for a \emph{direct} comparison with numerical simulations and is consistent with the conventions typically employed in chemical evolution studies. $[Z]^{\mathrm{phys}}$ can then be split into a young and old component as described in the following paragraphs. We will test the capabilities of this new definition in section~\ref{M83_test} with a 1-to-1 comparison to stellar probes.

To further analyze the chemical evolution history, we separate contributions from the young and old stellar populations. Following again \cite{Sextl2023,Sextl2024,Sextl2025M83} we introduce a conventional age boundary $t_{lim}^{y}$ to distinguish between populations dominated by recent star formation activity (young) and those tracing the longer-term assembly history (old). Thus, all SSPs with ages $t_{i} < t_{lim}^{y}$ are assigned to the young component, while those with $t_{i} \geq t_{lim}^{y}$ comprise the old component. We will use $t_{lim}^{y}$ = 0.1 (for M83) and 0.3 Gyr (for the other galaxies), in the next sections to distinguish between the young and old population. These age thresholds are established after reviewing the selected SSP templates of all bins to ensure the cutoff is meaningful. The earlier age split in galaxy M83 is consistent with our work on M83 \citep{Sextl2025M83} and necessary for a proper comparison with appendix~\ref{appendix}.

Metallicities and ages for the young and old populations are then calculated with the above equations, but the sums are carried out only over the old or young SSP, respectively. This separation between the old and young population is used for the calculation of ages and metallicities. The contribution to the observed spectrum by the two populations is then described by the corresponding sums $b_{\mathrm{young}}$ and $b_{\mathrm{old}}$, where $b_{\mathrm{young}} + b_{\mathrm{old}} = 1$ by construction.

This division is also applied to the mass-weighted coefficients $\Tilde{b_{i}}$ as defined above. The fractional mass contributions of the two populations were then computed as
\begin{align} \label{eq:split}
\Tilde{b}_{\mathrm{young}} &= \sum_{t_{i}< t_{lim}^{y}} \Tilde{b_{i}}, \\
\Tilde{b}_{\mathrm{old}} &= \sum_{t_{i}\geq t_{lim}^{y}} \Tilde{b_{i}}.
\end{align}
This separation highlights the strong discrepancy that often arises between light- and mass-weighted quantities: while the integrated light can be dominated by a relatively small number of very luminous young stars, the bulk of the stellar mass generally resides in the older populations. 

\subsection{A side note on the SSP template set} \label{subsec:SSPset}

Apart from the new metallicity definition, we emphasize that, besides the uncertainties inherent to the fitting algorithm itself, full-spectral fitting lives and dies with the choice of templates. Each template set is characterized by the number of ages and metallicities covered as well as the ingredients it was calculated from. The computation of a simple stellar population spectrum then begins with an assumed star formation event in which all stars are formed simultaneously. The adopted stellar initial mass function (IMF) then dictates the relative weighting of stars of different masses. Individual stellar spectra are drawn from an empirical or theoretical stellar library and mapped onto stellar evolutionary tracks, which specify how stars of a given mass and metallicity evolve with time. By integrating all masses and evolutionary stages, one obtains the total flux of the population at a fixed age and metallicity. \\
We use for these calculations the program FSPS (version 3.2, Flexible Stellar Population Synthesis code, \citet{Conroy2010,Conroy20102}) which gives the user the choice from different stellar libraries, stellar evolution tracks, and IMFs. For our SSP set, we chose the MILES library \citep{MILES2006}, MESA stellar evolution isochrones \citep{Dotter2016,Choi2016} and
a Chabrier \citep{Chabrier2003} initial mass function. As the empirical MILES library has only a limited number of 
stars with T $> 9000$ K \citep{Martins2007}, it is crucial to add additional spectra of hot massive stars \citep{Eldridge2017}, Wolf-Rayet types \citep{Smith2002}, AGB- \citep{Lancon2000}, post-AGB \citep{Rauch2003} and carbon stars \citep{Aringer2009}. By extending MILES with libraries for hot and post-MS stars, we improve its applicability in the complex environments of nuclear rings. This set, called 'MILES-SSP' was successfully applied and tested in our analysis of the TYPHOON data of M83 and NGC1365 \citep{Sextl2025M83,Sextl2024}. 

In summary, we use a grid with 52 age and ten metallicity values (520 SSPs in total). The metallicities go from [Z] $=0.5$ dex in steps of $\Delta [Z] = 0.25$ dex down to [Z] = $-1.75$ dex.  The logarithmic age grid starts at $0.1$ Myr ages and goes up to 12.5 Gyr (see also Fig.~2 in \citet{Sextl2023} for an illustration of the grid).

As stated in \citet{Sextl2025M83}, roughly one third of the models are $20$ Myr and younger. This is important for the analysis of very young stars and clusters, which we can expect in circumnuclear star-forming rings (ages $<5$ Myr as shown in \citet{Prieto2019}). We intentionally do not remove young low-metallicity SSPs to adjust the fit to choose higher metallicities in young populations. We note that \citet{Freitas2023, SilvaLima2025} had to make such exclusions because some of the low metallicity MILES SSP templates were deemed unsafe by \citet{Vazdekis2015}.\\
In our work, we use MILES as one library among several others and find that the lowest [Z] value is not taken at all by our analysis algorithm and that all regular SF rings in our sample are indeed metal rich or close to solar at the present time. In Section \ref{youngtempl}, we further test the severe consequences of a smaller age grid, removing the youngest SSPs ($<6$ Myr and $<30$ Myr respectively) and running the fitting procedure again.

\begin{figure*}
       \medskip
	\center \includegraphics[width=0.7\textwidth]{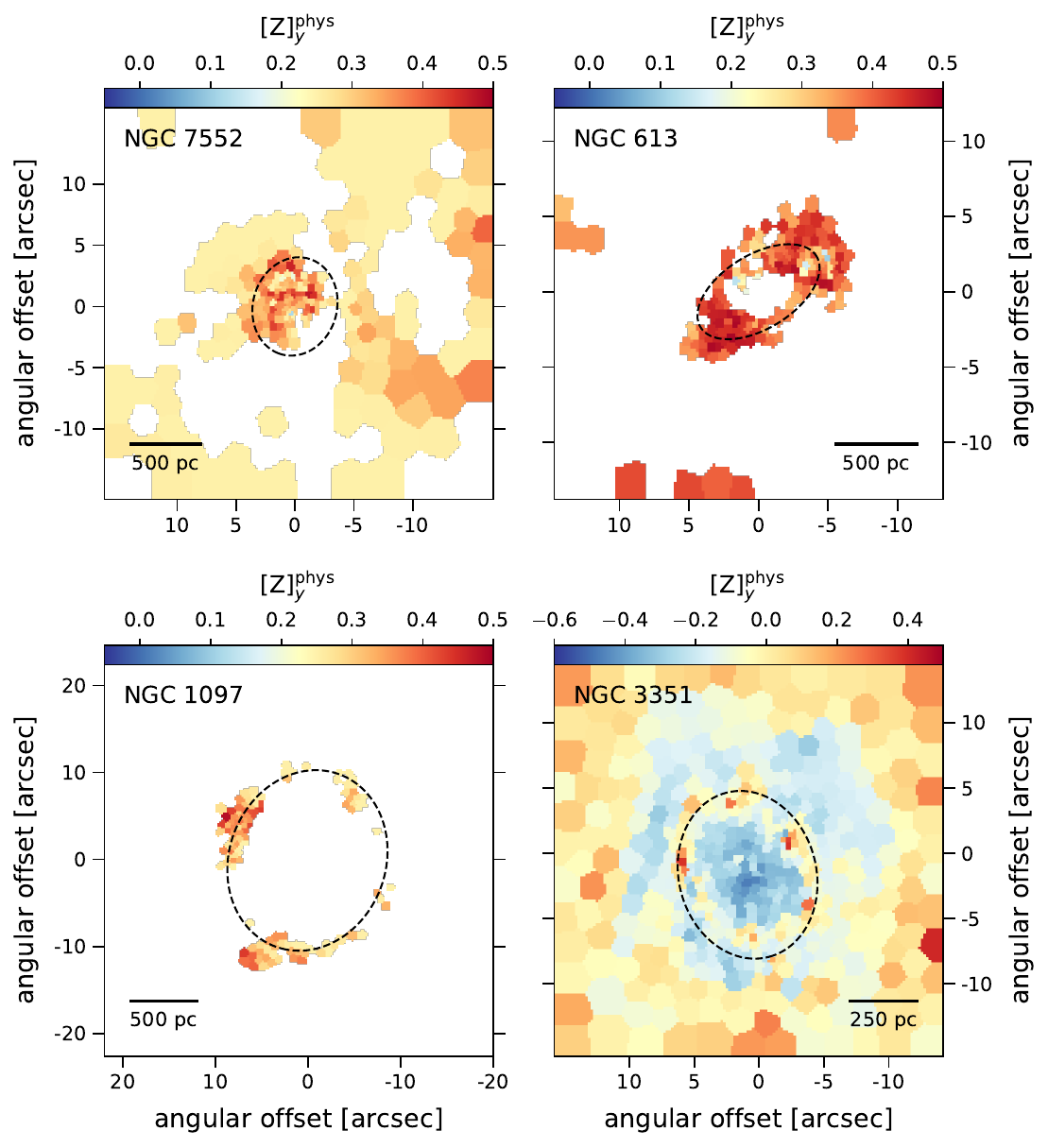}\medskip
    \center \includegraphics[width=0.98\textwidth]{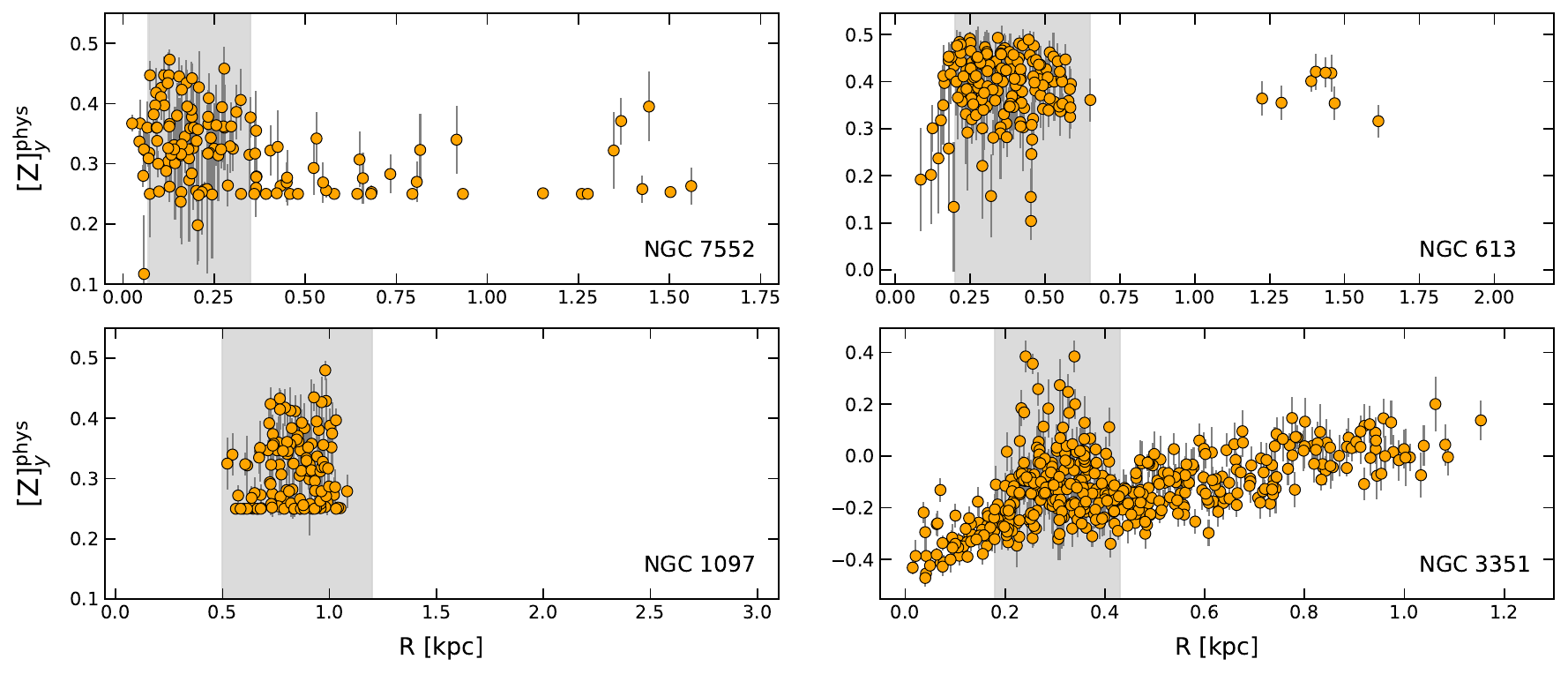}\medskip
	\caption{ Metallicity maps of the young stellar population (top) and radial metallicity distribution including uncertainties (bottom). }    \label{fig:Zyoung}
\end{figure*}

\begin{figure*}
       \medskip
	\center \includegraphics[width=0.9\textwidth]{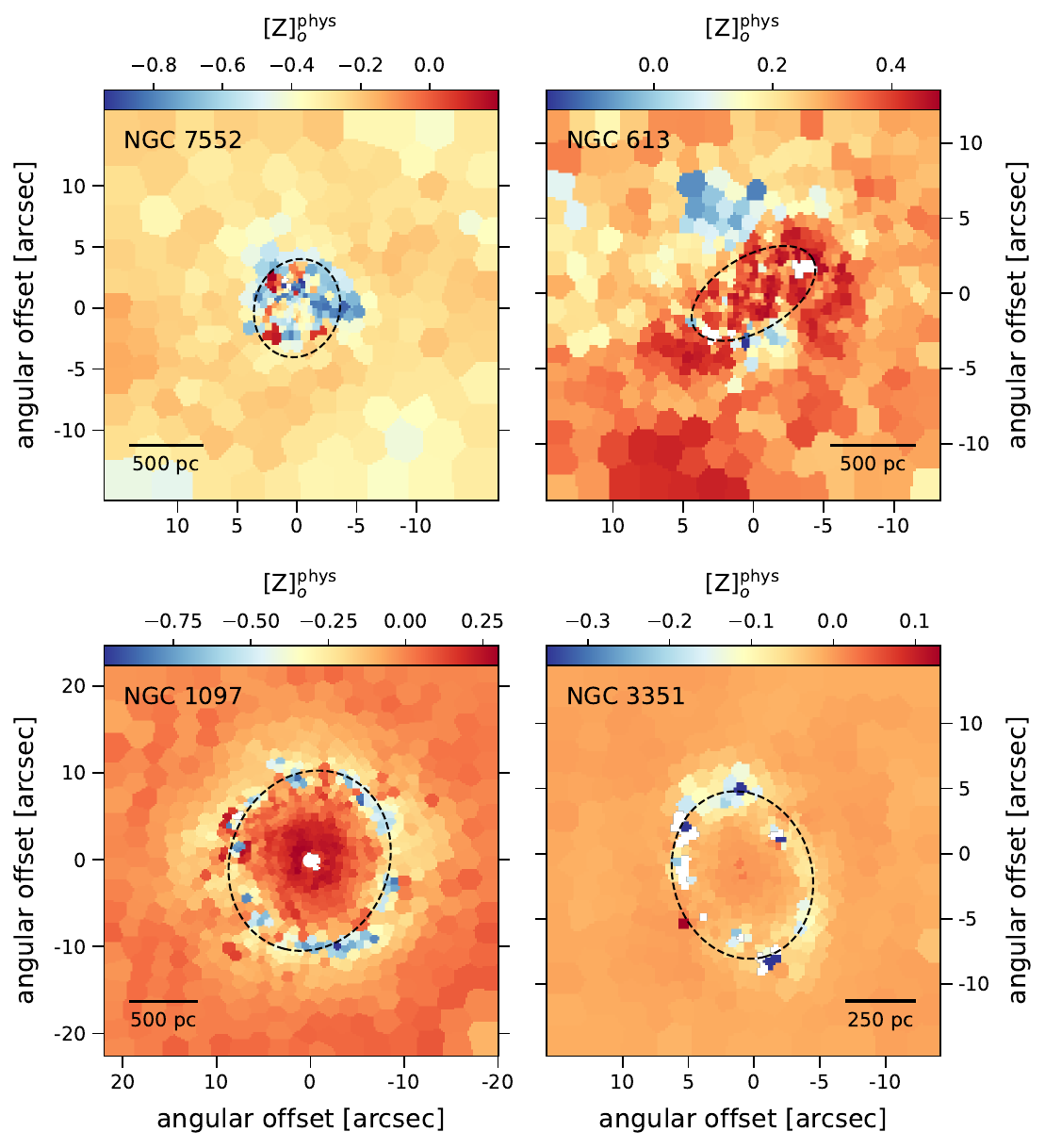}\medskip
	\caption{Metallicity maps of the old stellar population. For each galaxy, the approximate position of the nuclear ring is shown with a dashed line. Upper row: NGC 7552 (left), NGC 613 (right). Lower row: NGC 1097 (left), NGC 3351 (right)}    \label{fig:Zold_maps}
\end{figure*}

\begin{figure}
	\center \includegraphics[width=\linewidth]{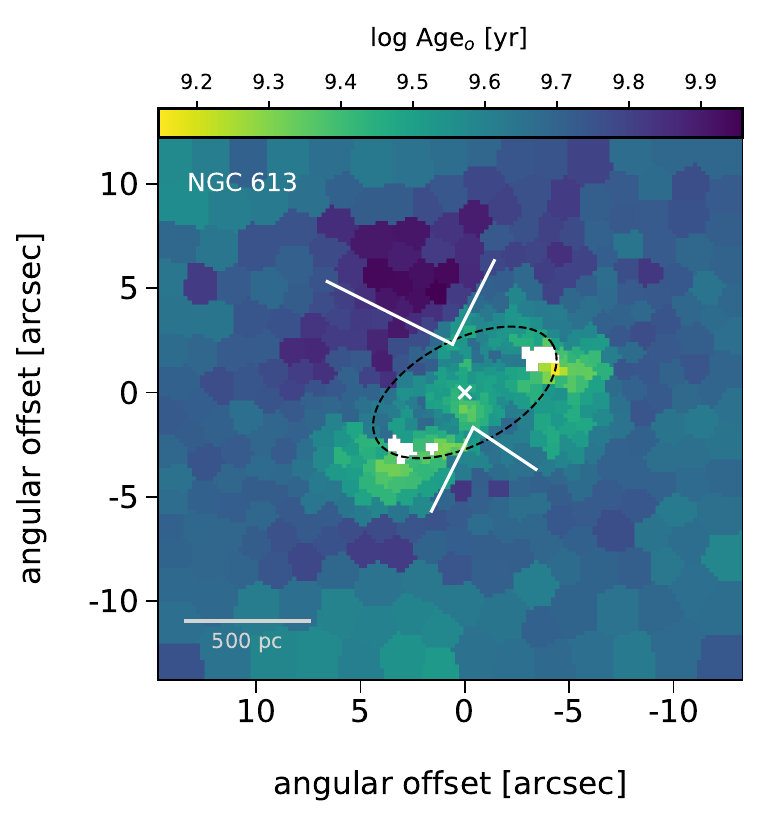}
	\caption{Map of the age of the old stellar population of NGC 613. The AGN outflow cone is indicated.}  \label{fig:AgeO_n613}
\end{figure}

\section{A MUSE metallicity test with young stellar probes in M83} \label{M83_test}

In addition to the galaxies in Table~\ref{tab:galaxies} MUSE observations are also available for the star forming galaxy M83. This galaxy harbors an asymmetrical nuclear region where H${\alpha}$ emission is irregularly distributed in the center \citep{Gadotti2019}. At a distance of 4.8 Mpc, its proximity makes M83 an excellent target for observations with the MUSE spectrograph, but it has also been a subject of accurate multi-object spectroscopic studies of individual stellar probes. \citet{Bresolin2016} measured the metallicities of individual blue supergiant stars throughout the disk of M83. In addition, young massive clusters in UV and optical (YMCs; \citet{Hernandez2018, Hernandez2019, Hernandez2021}) and super star clusters in the NIR (SSCs; \citet{Davies2017}) have been investigated spectroscopically. The results of this work offer the unique opportunity to compare the metallicities obtained from these stellar sources with those derived from our MUSE population synthesis analysis. \\
A similar comparison was carried out by \citet{Sextl2025M83} using TYPHOON data which have a lower spatial and spectral resolution (1.65 arcsec and 8\AA, respectively) but extend to shorter wavelengths (4000\AA). Here, we repeat the analysis employing the higher resolution of the MUSE-DEEP data cube. To improve the signal-to-noise ratio for spectral fitting, we combined individual spaxels within the vicinity of each stellar probe to produce integrated spectra representative of those regions. To reduce contamination from individual bright stars, we excluded spaxels within a radius of three pixels centered on each stellar probe. At the TYPHOON pixel scale of $1.65$" \citep{Grasha2022}, this adjustment was not required. However, with the much finer spatial resolution of MUSE at $0.2$", the light from a few bright stars can easily dominate the flux within a single pixel, and our fitting-procedure would lead to unreliable results. \\
The full spectral fitting was then applied to the binned spectrum to infer the stellar population properties surrounding the stellar probes, with particular emphasis on the metallicity $[Z]^{\mathrm{phys}}_{\mathrm{100 Myr}}$ of the young component (age $<100$ Myr). We subsequently compare this fitted metallicity with the independently measured metallicity obtained from the stellar probe. This comparison is quantified in terms of the difference $\Delta [Z] = [Z]_{\mathrm{probe}} - [Z]^{\mathrm{phys}}_{\mathrm{100 Myr}}$, which is illustrated in Figure \ref{fig:M83_compare}. Note that all stellar probes were adjusted to a common baseline of $Z_{\odot}=0.142$. Our results reveal a close agreement between the metallicity estimates derived from the full spectral fitting method and those from the stellar probe. The mean value of $\Delta [Z]$ is 0.04 dex and the scatter is 0.14 dex, the latter being the expected result for average errors of 0.1 dex for individual independent values of $[Z]_{\mathrm{probe}}$ and $[Z]^{\mathrm{phys}}_{\mathrm{100 Myr}}$, respectively.

Figure \ref{fig:M83_compare} demonstrates that MUSE IFU spectra are extremely useful to obtain quantitative information about the young stellar population despite their restricted wavelength range toward the blue. In \citet{Sextl2025M83} we argued in the opposite direction. However, this was based on the assumption of using low resolution TYPHOON spectra with moderate S/N. As we see now, the MUSE higher spectral resolution and high signal-to-noise ratio compensate for the lack of blue wavelength coverage.

We also carried out an FSF analysis of the MUSE-DEEP data of the central region of M83. The resulting map of $[Z]^{\mathrm{phys}}_{\mathrm{100 Myr}}$ is presented and discussed in the Appendix.

\begin{figure*}
       \medskip
	\center \includegraphics[width=0.99\textwidth]{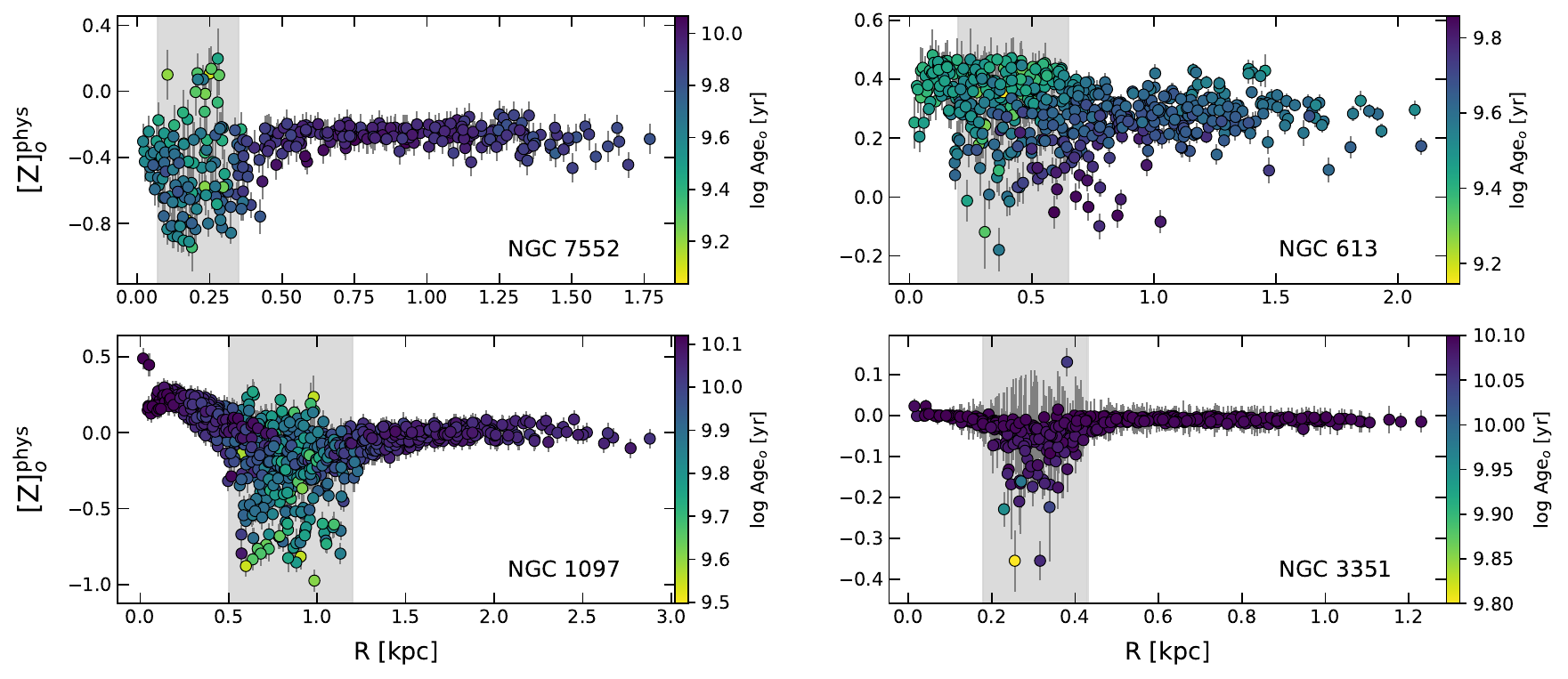}\medskip
	\caption{Radial metallicity of the old stellar population. Upper row: NGC 7552 (left), NGC 613 (right). Lower row: NGC 1097 (left), NGC 3351 (right). The position of the nuclear ring is again marked in gray. The color coding represents the values of log Age$_o$ [Gyr] which is different for each subfigure.}    \label{fig:Zold_grad}
\end{figure*}

\section{Ages and Extinction} \label{ages}

We start with the mean light-weighted ages in our sample (second column in Fig.~\ref{fig:Fig1_overview}). In each system except NGC 7552, the youngest stellar populations trace a clear ring-like morphology, consistent with the location of the nuclear ring inferred from both photometric and spectral diagnostics in the literature. In all four galaxies the contrast between the young star-forming ring and the exterior and interior stellar population is particularly sharp, confirming that these structures are relatively confined in both spatial extent and evolutionary timescale. In NGC 3351, a closer inspection of the regions with $\log {\rm Age} < 7$ reveals a direct correspondence with the prominent HST-identified clusters younger than 10 Myr \citep[][see also their Fig.~9]{Sun2024}. For NGC 613, the distribution of young stellar ages closely resembles the HST image presented in Fig.~1 of \citet{Barroso2014}. Similarly, in NGC 1097, our age map is consistent with the age of young clusters shown in Fig.~1a of \citet{Prieto2019}, while in NGC 7552 the general features are in line with those shown in Fig.~7 of \citet{Brandl2012}. \\

In the next section, we discuss stellar metallicity and distinguish between the young and old population. For this purpose, we use $t_{lim}^y$ = 0.3 Gyr as age boundary. Below and above this limit, we can calculate the mean ages of the young and old population. As it turns out, the mean ages of the young stars are much lower than the boundary, about 4 to 10 Myr in the nuclear star forming regions. Figure \ref{fig:AgeY_map} gives an example for NGC 3351. The locations of the three most prominent very young clusters are highlighted, indicating the presence of very young stars.

The average ages of the old stars are significantly higher and in the range of many Gyr as Figure \ref{fig:AgeO_grads} demonstrates. Notably, we observe a large dispersion of stellar ages within the nuclear star-forming regions, suggesting an extended period of star formation activity. This finding is consistent with previous studies by \citet{Allard2006, Sarzi2007,Knapen2008}, which present evidence that star formation in nuclear rings occurs episodically through multiple bursts, with activity sustained over long timescales rather than representing a one-time and short-lived event.

We also note an interesting bimodality with respect to the maximum stellar ages in Figure \ref{fig:AgeO_grads}. While for NGC 1097 and NGC 3351 the oldest stars in the nuclear rings have an almost similar age as out- and inside, they are significantly younger for NGC 7552 and NGC 613. It seems that in the latter two cases stars have started to form later or older stars have migrated away from the ring.

In FSF, we can also determine the interstellar extinction A$_V$ shown in Fig.~\ref{fig:Fig1_overview} (third row). The characteristic dust lanes flowing along the leading edges of the stellar bars are observed to connect directly to the nuclear ring, in agreement with predictions from dynamical models of bar-driven inflows \citep{Verwilghen2024}. These dust structures trace the flow of interstellar material from the larger bar region toward the circumnuclear zone, feeding ongoing star formation in the ring. Two aspects seem to be noteworthy: First, the regions of the most prominent cluster in the rings seem to show relatively little reddening and extinction. Figure \ref{fig:AgeY_map} gives an example for the case of NGC 3351. This was already shown in Fig.~24 in \citet{Emsellem2022}. However, the authors there speak of a technical artifact in the fit converging to an 'misleading local minimum' with very low $[Z]^{\mathrm{lw}}$, young age, and low E(B-V) values, possibly created from a lack of young templates. We will show in the next section that the physical metallicity of the young stars in these regions is close to solar. At the same time, all of our fitting runs indicate indeed low extinction within these regions. From a stellar evolutionary standpoint, this finding is easily explained by the onset of stellar winds from newly formed stars, which effectively disperse the residual parental material from which the clusters originated. \citet{Prieto2019} extracted the gas extinction from the HST recombination map $H{\alpha} / P{\alpha}$ of NGC 1097 and came to the same conclusion. Clusters as young as $4$ Myr have effectively removed dust from their surroundings. \citet{Sun2024} also reported YMCs that lose their local gas and dust reservoirs at ages between $3$ and $6$ Myr in NGC 3351. \citet{Knutas2025} found similar timescales in M83 using JWST observations. \\
As a second peculiar feature, on the rings and inside toward the very nucleus, we consistently detect a region largely free of dust attenuation. The central dust-deficient zone is ubiquitous across the sample, suggesting that the innermost few hundred parsecs are not efficient reservoirs of cold material. This is also the case in NGC1365 \citep{Sextl2024}. For M83, however, a dust cavity is found on the star forming ring (see \citealt{Sextl2025M83}), while the rest of the central region shows increased interstellar reddening.

\begin{figure*}
       \medskip
	\center \includegraphics[width=0.9\textwidth]{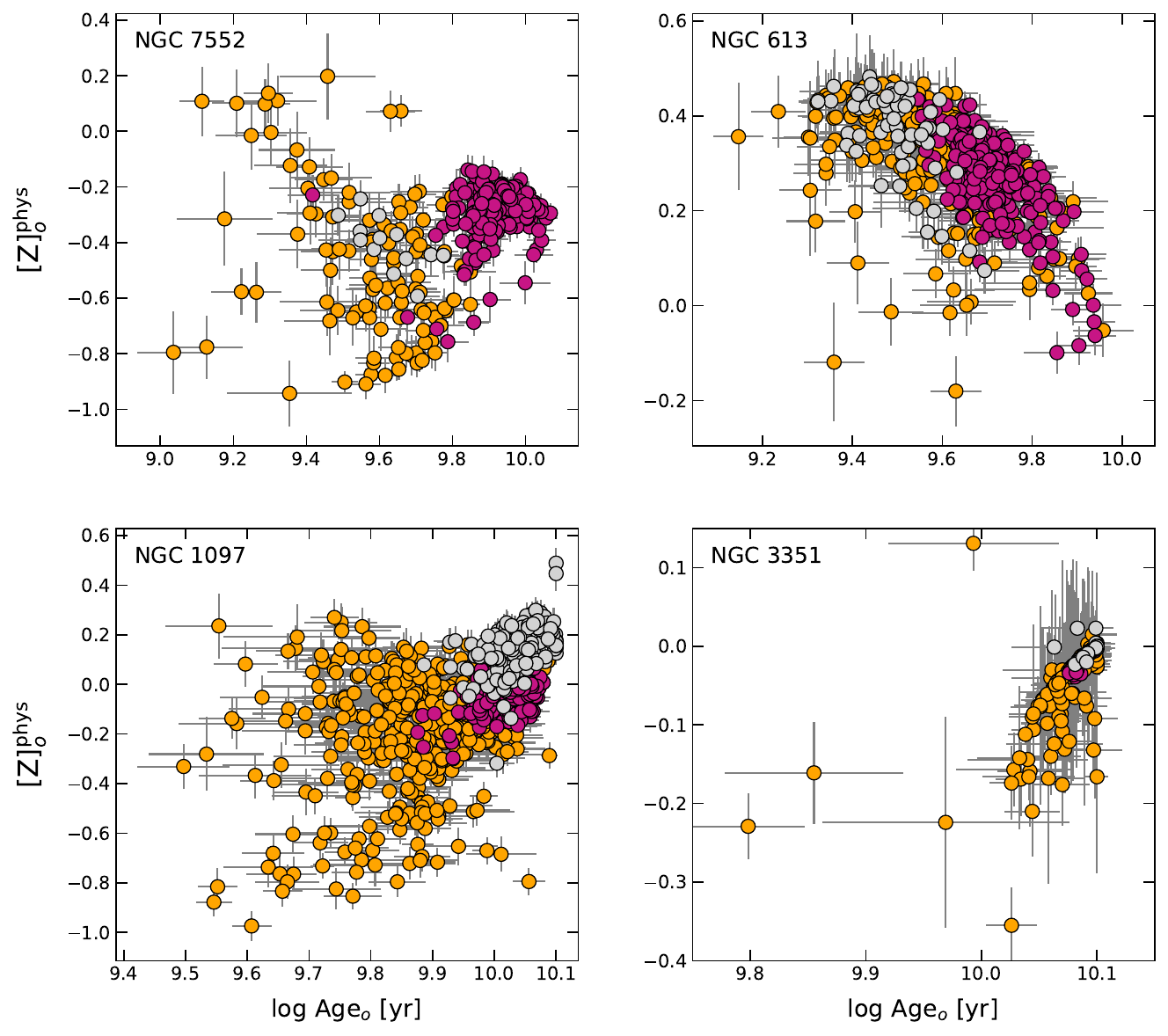}\medskip
	\caption{Metallicity versus age of the old stellar population. Voronoi bins outside the rings appear in violet red, those on the ring are colored orange, and inner sections are light gray. Upper row: NGC 7552 (left), NGC 613 (right). Lower row: NGC 1097 (left), NGC 3351 (right).}    \label{fig:Zold_age}
\end{figure*}

\begin{figure*}
       \medskip
	\center \includegraphics[width=0.99\textwidth]{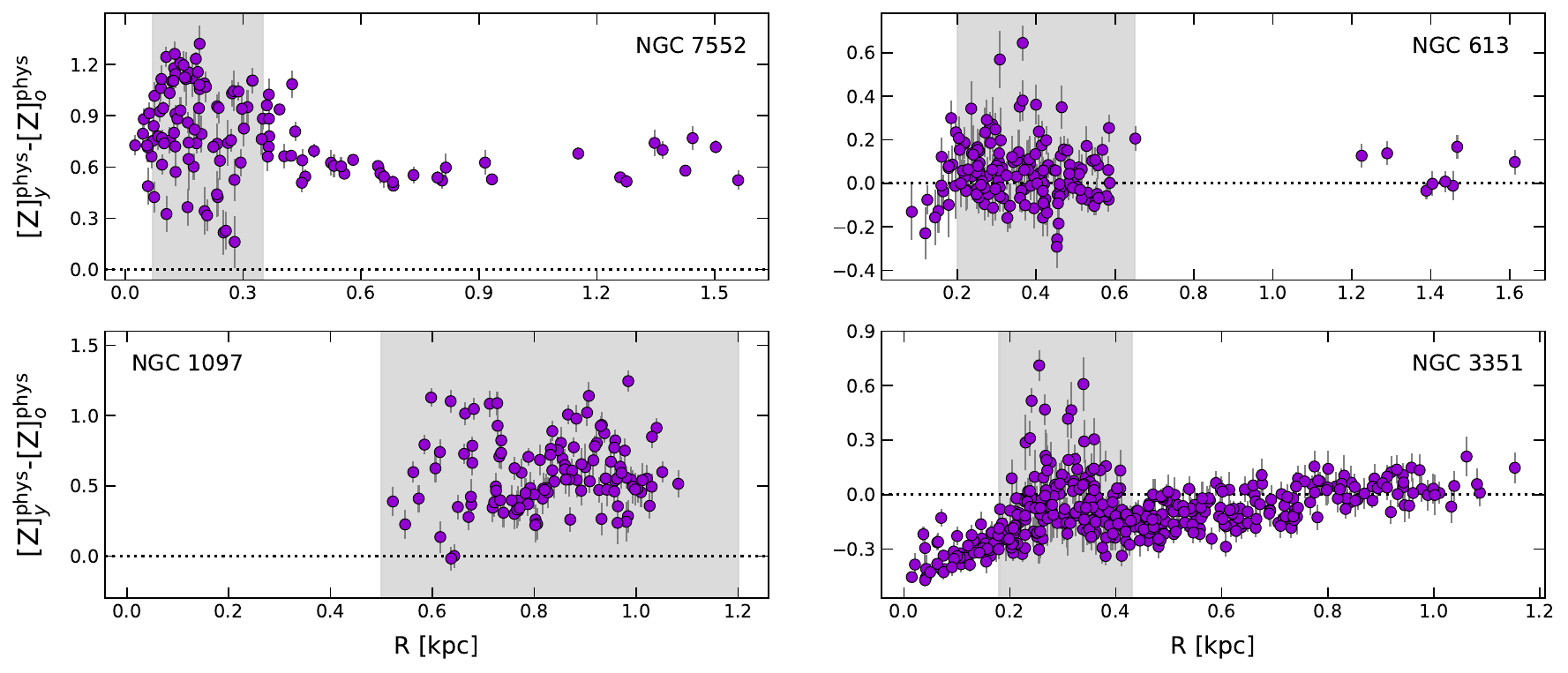}\medskip
	\caption{Metallicity difference [Z]$_y^{phys}$ - [Z]$_o^{phys}$ versus galactocentric distance. Upper row: NGC 7552 (left), NGC 613 (right). Lower row: NGC 1097 (left), NGC 3351 (right). The dashed line marks the zero line.}   \label{fig:Zdiff}
\end{figure*}

\begin{figure} 
	\center \includegraphics[width=0.86\linewidth]{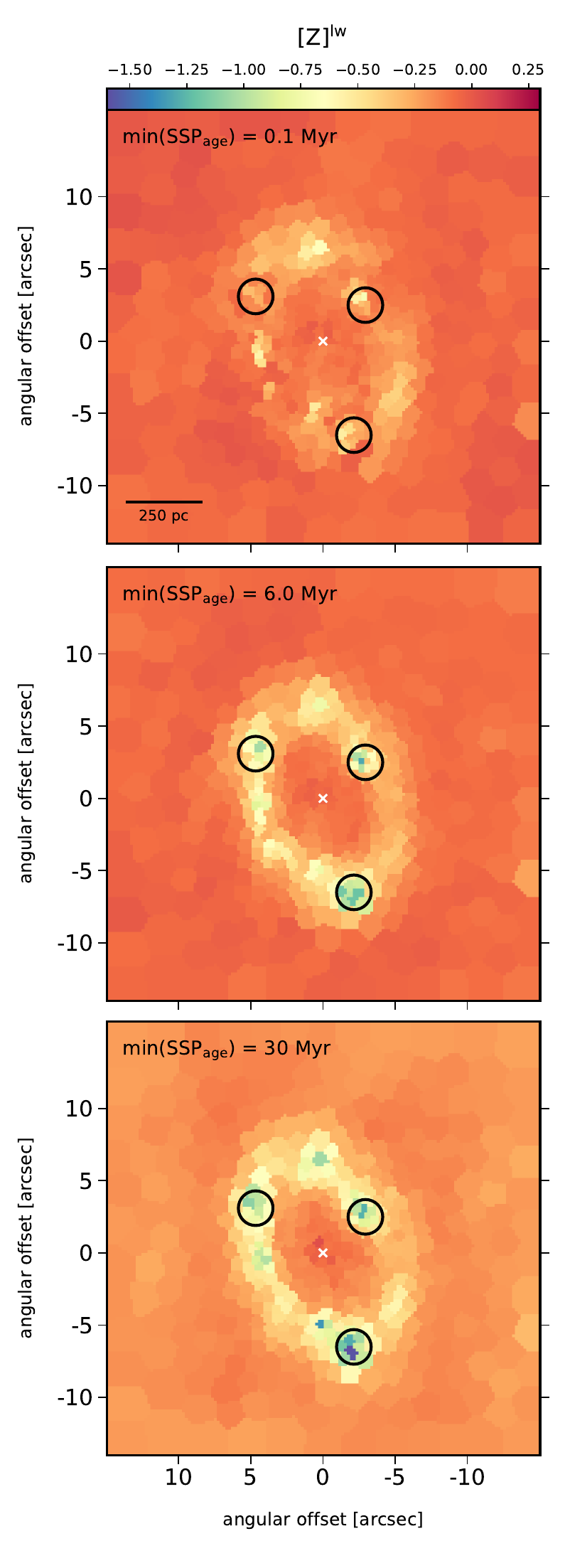}
	\caption{Map of total light-weighted metallicity $[Z]^{\mathrm{lw}}$ in central region of NGC 3351 obtained with SSP sets of different age ranges: (top) 0.1 Myr to 12.5 Gyr; (middle) 6 Myr to 12.6 Gyr; (bottom) 30 Myr to 12.6 Gyr. The color bar is identical for all subplots. See text.}  \label{fig:Zlw_SSP}
\end{figure}

\begin{figure} 
	\center \includegraphics[width=0.86\linewidth]{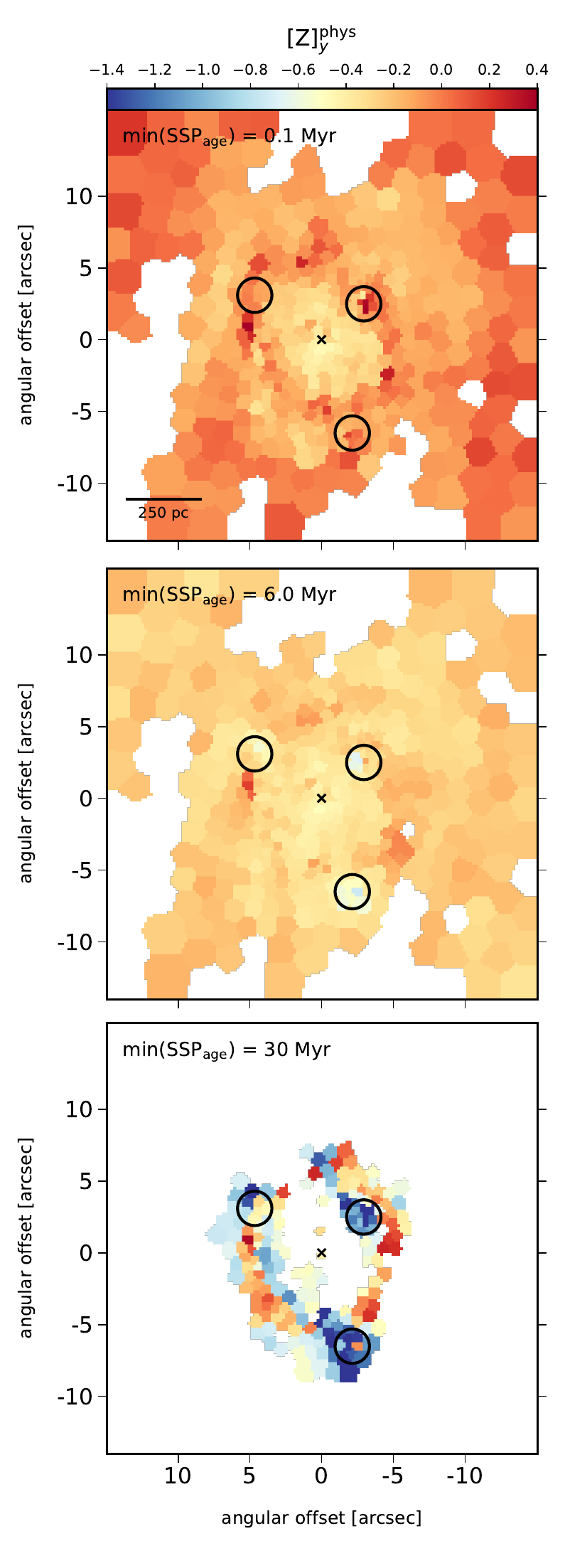}
	\caption{Map of physical metallicity $[Z]^{phys}_{y}$ in central region of NGC 3351 obtained with SSP sets of different age ranges: (top) 0.1 Myr to 12.5 Gyr; (middle) 6 Myr to 12.5 Gyr; (bottom) 30 Myr to 12.5 Gyr. The color bar is identical for all subplots. See text.}  \label{fig:Zyoung_SSP}
\end{figure}

\section{The metallicity of nuclear star forming regions}
The fourth row of Fig.~\ref{fig:Fig1_overview} shows maps of the overall light-weight metallicity $[Z]^{\mathrm{lw}}$. We find peculiar drops of $[Z]^{\mathrm{lw}}$ in our sample, qualitatively confirming the results of \citet{Pessa2023}, \citet{Emsellem2022}, \citet{Rosado2020}, \citet{Navarro2024} and \citet{Bittner2020}, although our low values of $[Z]^{\mathrm{lw}}$ are (partially) less extreme. Even after the inclusion of our very young stellar templates and a nebular continuum correction (see \citealt{Sextl2024,Sextl2025M83}), the maps show these features. This was already discussed in the appendix of \citet{Pessa2023} and we will extend the discussion in Section~\ref{youngtempl}. 

However, we need to keep in mind that what we see here is a mean quantity summarized over an enormously wide range of stellar ages and weighted by stellar light. It is important to understand that the low $[Z]^{\mathrm{lw}}$ values found in regions considered 'young' do not directly reflect the intrinsic metallicity of the youngest stars. In such regions often one third or even half of the light still comes from older stellar populations.

Therefore, we calculate [Z] using our physically motivated definition and divide between the young ($\leq 0.3$ Gyr) and the old population, which significantly alters the picture. Maps of the young stellar population $[Z]^{\mathrm{phys}}_y$ are provided for each galaxy in Fig.~\ref{fig:Zyoung}. Only regions with a sufficient light fraction of young stars ($b_y>0.2$) are shown. Voronoi bins that do not meet this criterion are left blank. In addition, Figure ~\ref{fig:Zyoung} shows the radial galactocentric distribution of $[Z]^{\mathrm{phys}}_y$.

For NGC 1097, NGC 613, NGC 7752 the young component does not fall below solar metallicity. In contrast, many regions show clear super-solar enrichment, with $[Z]^{\mathrm{phys}}_y \gtrsim 0.2$ dex.

For NGC 7552, we note that this is consistent with the findings reported by \citet{Calzetti2010, Moustakas2010, Wood2015}, who despite some ambiguities in the gas-phase metallicity measurements, classify the galaxy as at least solar and likely super-solar. 

In NGC 3351, the metallicity of the young population is significantly lower than in the other three galaxies, but we also see clear enrichment in the circumnuclear ring. The regions of the three most prominent young clusters discussed before show solar or somewhat higher metallicity. Generally, the young star metallicity reaches a pronounced local maximum in the ring of NGC 3351. We also note the very clear drop inside the ring where $[Z]^{\mathrm{phys}}_y$ is 0.2 dex lower than outside. \citet{Diaz2007} were able to derive direct HII region abundances on the star forming ring of this galaxy. The authors obtained metallicity values ranging from [Z]=$-0.27 \pm 0.11$ (when using \citealt{Asplund2009} as the solar standard) in the eastern segment of the ring up to $0.08 \pm0.09$ in the north-western and southern parts. This aligns well with our results for $[Z]^{\mathrm{phys}}_y$ spreading from $-0.3\pm 0.1$ dex up to $0.3\pm0.1$ dex, especially considering the potential effects of dust depletion of oxygen in HII-regions \citep{Bresolin2025}.\\

The physical metallicity $[Z]^{\mathrm{phys}}_o$ of the older stellar population ($\geq 0.3$ Gyr) is shown in Figures \ref{fig:Zold_maps} (maps) and \ref{fig:Zold_grad} (galactocentric distribution). We find an extremely wide range of $[Z]^{\mathrm{phys}}_o$ in the nuclear star forming regions of all four galaxies. Outside these regions, the dispersion of metallicity is significantly smaller. We encounter very low metallicities in the rings of NGC 1097 and NGC 7552.

While the dispersion of $[Z]^{\mathrm{phys}}_o$ outside the nuclear rings is small, the values differ between the four galaxies. $[Z]^{\mathrm{phys}}_o$ is clearly super-solar in NGC 613, solar in NGC 1097 and NGC 3351 and a factor of two below solar in NGC 7552. This must be the result of different evolution histories outside the nuclear rings.

In the case of NGC 613 we notice areas of lower $[Z]^{\mathrm{phys}}_o$ in projected regions perpendicular to the nuclear ring. As Figure \ref{fig:AgeO_n613} indicates, these stars have the highest age within the old population encountered in this galaxy (see also Figure \ref{fig:Zold_age}). These areas seem to coincide with the AGN outflow cone \citep{SilvaLima2025}. Focusing only on the metallicity of the older population in the light-weighted analysis, or considering mean quantities without applying an age cut, does not reveal this distinctive feature. From our viewpoint, the northern cone is projected in front of the stellar ring and is therefore more accessible, while the southern cone lies behind the ring. However, the stars inside the northern cone can only be clearly identified farther out, where the bright stellar ring and its long history of star formation no longer dominates the view. We speculate that the outflow has interrupted star formation for a long period of time and no substantial chemical enrichment was possible. Similar IFU evidence for small-scale negative feedback was reported by \citet{Feuillet2025} in the intermediate spiral galaxy NGC 7469. 

Figure \ref{fig:Zold_age} shows $[Z]^{\mathrm{phys}}_o$ as a function of the average age of the old stars in and outside the nuclear star forming regions. The presence of old stars born with very low metallicity over many Gigayears in the cases of NGC 1097 and NGC 7552 is indicative of infall of low metallicity gas over a long period of time. For NGC 613, on the other hand, we see the signature of normal chemical evolution less affected by infall. The old population in the central region of NGC 3351 that we see now was mostly born more than $10$ Gyrs ago.

A coarse look at the chemical evolution history is also provided by Figure \ref{fig:Zdiff}, where we plot the metallicity difference between the young and old population as a function of galactocentric distance. The positive differences indicate normal chemical evolution with the scatter induced by the effects of infall of metal-poor gas of different strengths. NGC 3351, on the other hand, the mostly negative values in the center hint at the most recent strong infall of metal-poor gas. 

\section{Importance of a young template grid} \label{youngtempl}

In the sections above, we have pointed out several times that it is crucially important to include the SSP contributions of very young stars with a wide range of possible metallicities. Due to its importance, we use this additional section to address the issue in more detail. We select NGC 3351 and its nuclear ring as an example.

We compare the results of the FSF analysis of the central region of NGC 3351 obtained with three sets of SSPs. The first is identical to the one described in Section \ref{subsec:SSPset}. The other two use the same range of metallicities but the ages range from $6$ Myr to $12.5$ Gyr for set two and $30$ Myr to $12.5$ Gyr for set three, respectively. For the ages included, the age steps are the same in all three sets.

Figure \ref{fig:Zlw_SSP} shows the comparison of the total light-weighted metallicities $[Z]^{\mathrm{lw}}$. As discussed in Section 7, we encounter spurious regions of low metallicity on the nuclear ring as an artifact of the luminosity-weighted mean over all ages, young and old. However, now we also see the dramatic influence of the contribution of the youngest stars in the sets of SSP. With stars younger than $6$ Myr omitted, the metallicities drop to values as low as $[Z]^{\mathrm{lw}}$ = $-1.2$. The effect becomes even more extreme when SSP younger than $30$ Myr are left out. We encounter extreme regions with $[Z]^{\mathrm{lw}}$ = $-1.7$.

The effects shown in Figure \ref{fig:Zlw_SSP} were already partially discussed in \citet{Pessa2023}. We have included the figure here because of the ongoing discussion about the chemical evolution of nuclear rings. However, as we have explained in Section 4.2 the physical relevant description of stellar metallicity is not given by $[Z]^{\mathrm{lw}}$ but rather by $[Z]^{\mathrm{phys}}$ as defined by equations (5) and (6). Thus, it is important to investigate how the physical metallicities of the young and old stellar populations are affected by the choice of SSPs. Figure \ref{fig:Zyoung_SSP} displays maps of the physical metallicity $[Z]^{\mathrm{phys}}_y$ of the young population for the three cases. The differences are equally dramatic, and it is obvious that leaving out the contribution of the youngest stars can cause significant systematic effects. For example, physical metallicities are on average $0.2$ dex lower when stars younger than $6$ Myr are not included. Differences larger than $1.0$ dex are encountered when the contributions of stars younger than 30 Myr are neglected. We have also tested the influence of the SSP sets on the determination of the old population physical metallicities $[Z]^{\mathrm{phys}}_o$. We find that the systematic effects are small.

\section{Summary and conclusions} \label{conclusions}

In our stellar population synthesis of the nuclear star forming regions of four galaxies we use a physical definition of stellar metallicity  which is consistent with the study of chemical evolution. In addition, we disentangle the contributions of young ($\leq 0.3$ Gyr) and old stars. In this way, we avoid methodological artifacts resulting from the use of conventional luminosity- or mass-weighted averages over stars of all ages, young and old. We also demonstrate that it is crucially important to include the contributions of very young stars in the analysis.

We conclude that each galaxy in this study has its own unique characteristics shaped by its individual star formation history, despite sharing a barred morphology and similar stellar mass. Our spectral analysis reveals that the stellar populations currently forming in the nuclear rings in NGC 613, NGC 1097, and NGC 7752 are super-solar. On the ring of NGC 3351 the metallicities are in a range between half and twice solar. This is in agreement with direct HII region abundance measurements. We also see a clear metallicity enrichment in the case of the ring of NGC 3351. The metallicity distribution in the nuclear ring is similar to the 'onfall' scenario discussed in \citet{Friske2023}, Fig.~6. 

The ages and metallicities of the old stars indicate continuous star formation in the presence of an inflow of low metallicity gas over many Gyrs in the case of NGC 1097 and NGC 7752. For NGC 613, low metallicity infall appears to be less important. In the case of NGC 3351, the old stars were generated mostly ten Gyrs ago. In the very center, the lower metallicity of the young stars indicates the most recent strong infall of lower-metallicity gas.

The infall scenario is supported by the reddening maps obtained with our population synthesis technique. We find dust lanes tracing the flow of interstellar material toward the circumnuclear zone and providing the material for ongoing star formation. The region inside the nuclear rings is largely free of dust attenuation. On the rings, prominent stellar clusters show little extinction, very likely as a result of the onset of strong stellar winds. 

After extensive work on several barred spiral galaxies, which include the four listed in this work, M83/NGC5236 \citep{Sextl2025M83} as well as NGC 1365 \citep{Sextl2024}, we can cautiously draw an unified picture. The earliest stars formed after a substantial gas inflow in these rings tend to be relatively metal-poor (as low as [Z] $\sim – 0.9$ in NGC 7552, Figure~\ref{fig:Zold_age}), reflecting the primordial composition of gas coming from the outer disk or from galaxy interactions. Later generations within these star-forming rings show normal gradual chemical enrichment, eventually matching or surpassing solar metallicity. In some galaxies, such as NGC 7552 and NGC 1097, additional inflow events occurred over the past billion years, resulting in stellar populations that reflect both new inflow and ongoing chemical evolution. In a third case, galaxies like M83 and NGC 1365, the most recent gas inflow occurred only a few hundred million years ago. It is not surprising that young, low-metallicity stars dominate the spectra of these central regions.

In summary, we conclude that the high spatial and spectral resolution of the MUSE IFU spectrograph combined with the power of an 8m VLT mirror and our technique of stellar population synthesis provide unique means to investigate the nuanced nuclear star forming regions of galaxies in the nearby universe.
 
\begin{acknowledgments}
Acknowledgments. We thank our referee for the careful review of our manuscript, and we are grateful to Dimitri Gadotti and Svea Hernandez for constructive and critical discussion and valuable comments during the course of this work. We acknowledge support from the Munich Excellence Cluster Origins and the Munich Institute for Astro-, Particle, and Biophysics (MIAPbP) both funded by the Deutsche Forschungsgemeinschaft (DFG, German Research Foundation) under Germany's Excellence Strategy EXC-2094 390783311. In addition, ES has been supported by the European Research Councel COMPLEX project under the European Union's Horizon 2020 research and innovation program grant agreement ERC-2029-AdG 882679.
Based on observations collected at the European Organisation for Astronomical Research in the Southern Hemisphere under ESO programme 096.B-0057, and processed data created thereof. Additionally based on data obtained from the ESO Science Archive Facility with DOI \url{https://doi.org/10.18727/archive/42}. 
\end{acknowledgments}

\facilities{ESO VLT/MUSE}

\appendix
\renewcommand\thefigure{\thesection.\arabic{figure}}    
\setcounter{figure}{0}    

\section{The center of M83 with MUSE} \label{appendix}

Our work focuses on the properties of nuclear rings in the galaxies NGC 7552, NGC 613, NGC 1097, and NGC 3351. In these systems, the young stellar component generally shows metallicities two to three times higher than solar, except in NGC 3351, where values range from approximately half to twice solar. No regions with lower metallicities than this were encountered. That raises a more general question: Does this trend more or less hold for all star-forming nuclear rings, or do individual other cases differ and how are the metallicities distributed?

In our earlier work \citep{Sextl2025M83}, we investigated the circumnuclear ring of M83 reconstructed from CO maps shown in \citet{Harada2019}. We reported a coherent drop in the metallicity of the young stellar population along the southern part of the ring, based on TYPHOON survey data. The IFU-like TYPHOON observations have a coarser spatial and spectral resolution ($1.65$" per spaxel, 8 \AA) but extend further into the blue wavelength regime down to 4000\AA\, \citep{Grasha2023,Chen2023}. Thus, a comparison with an analysis based on MUSE data is very interesting.\\
The central region of M83 has also been observed within MUSE-DEEP, offering a rare opportunity to compare full-spectral fitting results from two distinct observational campaigns. For this purpose, we again applied our pPXF  pipeline now to MUSE data of the central region of M83 and chose our standard template set as described in section~\ref{subsec:SSPset}. 

\begin{figure}  
	\center \includegraphics[width=0.55\linewidth]{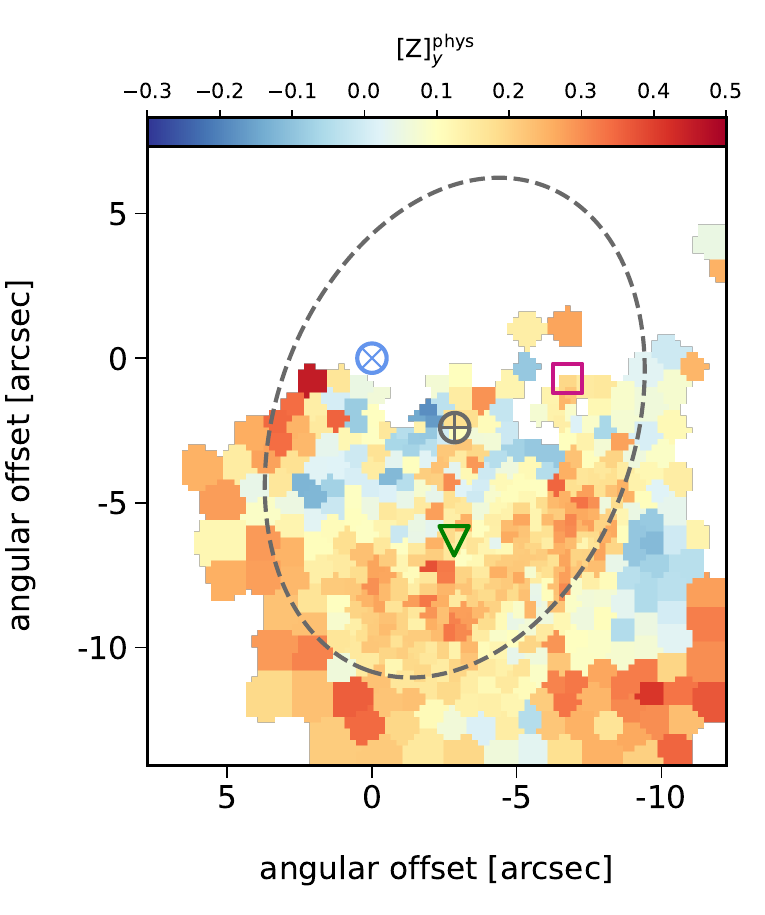}
        \begin{minipage}{0.75\linewidth}
        \caption{
            Metallicity of the young stellar population in the center of M83 with MUSE. Only Voronoi bins with a sufficient young stellar component ($b_y>0.2$) are shown. The nuclear ring by \citet{Harada2019} reconstructed from CO observations is drawn as dashed ellipse. Its center is determined from the CO velocity field \citep{Muraoka2009} and shown with a grey plus. The green triangle shows the potential AGN position (pointing P3) from \citet{Hernandez2025} (RA = $204.2530486$, Dec = $-29.867170$, (J2000), priv. comm.). The optical nucleus \citep{Thatte2000} is shown as a blue cross and the stellar kinematic center (Model A) is shown as violet square \citep{Della_Bruna2022}.}  \label{fig:M83_zy}
        \end{minipage}
\end{figure}

Figure~\ref{fig:M83_zy} shows the metallicity distribution of stars younger than $100$ Myr, derived from Voronoi spectra with a S/N of $250$. This can be directly compared with \citet{Sextl2025M83} Figure~17, based on TYPHOON. We again detect the metallicity drop along the southwestern portion of the ring, as well as the slightly sub-solar values west of the optical center. In addition, certain regions east of the center appear slightly metal-poorer in MUSE than in TYPHOON ($\sim 0.25$ dex difference). \\

These results demonstrate that our revised metallicity definition, combined with a consistent fitting approach, yields results that are broadly comparable across different observational campaigns with different spectrographs, different wavelength regimes, and varying spatial and spectral resolution. They also hint that not all nuclear star-forming rings share the same morphological or chemical characteristics. M83 is distinct in the spatial distribution of the lower metallicity bins. 
They are not regularly distributed along the ring (regions with $[Z]_y^{\mathrm{phys}}>0$ dex are found in the south-east). Neither do the regions within the ring appear regular. We find a coherent area in the south, which is relatively enriched, and an irregular distribution around the nucleus. \\
This fits well with M83's highly peculiar kinematic properties \citep{Della_Bruna2022}. Its dynamical center does not coincide with the optical center \citep{Thatte2000,Diaz2006}, and the location of its elusive AGN remains uncertain \citep{Hernandez2025}. 
Together, these findings highlight the need to examine galaxies individually, as the metallicity and stellar age distribution within a nuclear ring are shaped by the unique evolutionary history of its host galaxy.

\bibliography{Nuclear_Rings.bib}{}
\bibliographystyle{aasjournalv7}

\end{document}